\documentclass[prl,twocolumn,showpacs,preprintnumbers,amsmath,amssymb]{revtex4}
\usepackage{graphicx}
\usepackage{dcolumn,color}
\usepackage{graphics}
\usepackage{ifpdf,latexsym,overpic,rotating}
\usepackage{bm}
\usepackage{multirow}
\usepackage{epsfig}
\usepackage[english]{babel}

\begin{document}
\bibliographystyle{try}
\topmargin 0.1cm
\newcommand{\bc}           {\begin{center}}
\newcommand{\ec}           {\end{center}}
\newcommand{\bq}           {\begin{eqnarray}}
\newcommand{\eq}           {\end{eqnarray}}
\newcommand{\be}           {\begin{equation}}
\newcommand{\ee}           {\end{equation}}
\newcommand{\bi}           {\begin{itemize}}
\newcommand{\ei}           {\end{itemize}}
\newcommand{\er}{$\pm$}

\newcommand{\DpiL}{$\Delta\pi_{L<J}$}
\newcommand{\DpiH}{$\Delta\pi_{L>J}$}
\newcommand{\PIIpi}{$P_{11}(1710)\pi$}
\newcommand{\FISpi}{$F_{15}(1680)\pi$}
\newcommand{\Roppi}{$P_{11}(1440)\pi$}
\newcommand{\DIzpi}{$D_{13}(1520)\pi$}
\newcommand{\SIIpi}{$S_{11}(1535)\pi$}
\newcommand{\Deta}{$\Delta\eta$}

\newcounter{univ_counter}
\setcounter{univ_counter} {0} 
\addtocounter{univ_counter} {1}
\edef\HISKP{$^{\arabic{univ_counter}}$}
\addtocounter{univ_counter}{1}
\edef\GIESSEN{$^{\arabic{univ_counter}}$}
\addtocounter{univ_counter}{1}
\edef\GATCHINA{$^{\arabic{univ_counter}}$}
\addtocounter{univ_counter}{1} 
\edef\KVI{$^{\arabic{univ_counter}}$} 
\addtocounter{univ_counter}{1} 
\edef\PI{$^{\arabic{univ_counter}}$} 
\addtocounter{univ_counter}{1} 
\edef\FSU{$^{\arabic{univ_counter}}$}
\addtocounter{univ_counter}{1}
\edef\BASEL{$^{\arabic{univ_counter}}$}
\addtocounter{univ_counter}{1}

\title{\boldmath Three-body nature of $N^{\bf *}$ and $\Delta^*$ resonances from sequential decay chains
}
\author{
\renewcommand{\thefootnote}{\alph{footnote}}
A.~Thiel\HISKP,
V.~Sokhoyan\HISKP,
E.~Gutz\HISKP$^,$\GIESSEN,
H.~van~Pee\HISKP,
A.V.~Anisovich\HISKP$^,$\GATCHINA,
J.C.S.~Bacelar\KVI,
B.~Bantes\PI,
O.~Bartholomy\HISKP, 
D.~Bayadilov\HISKP$^,$\GATCHINA,
R.~Beck\HISKP,
Yu.~Beloglazov\mbox{\GATCHINA,}
R.~Castelijns\KVI,
V.~Crede\FSU,
H.~Dutz\PI,
D.~Elsner\PI,
R.~Ewald\PI,
F.~Frommberger\PI,
M.~Fuchs\HISKP,
Ch.~Funke\HISKP,
R.~Gregor\GIESSEN,
A.~Gridnev\mbox{\GATCHINA,}
W.~Hillert\PI,
Ph.~Hoffmeister\HISKP,
I.~Horn\HISKP,
I.~Jaegle\BASEL,
J.~Junkersfeld\HISKP,
H.~Kalinowsky\mbox{\HISKP,}
S.~Kammer\PI,
V.~Kleber\PI,
Frank~Klein\PI,
Friedrich~Klein\PI,
E.~Klempt\HISKP,
M.~Kotulla\GIESSEN$^,$\BASEL,
B.~Krusche\BASEL,
M.~Lang\HISKP,
\mbox{H.~L\"ohner\KVI,}
\mbox{I.~Lopatin\GATCHINA,}
S.~Lugert\GIESSEN,
T.~Mertens\BASEL,
 J.G.~Messchendorp\KVI,
V.~Metag\GIESSEN,
B.~Metsch\HISKP,
\mbox{M.~Nanova\GIESSEN,}
\mbox{V.~Nikonov\HISKP$^,$\GATCHINA,}
D.~Novinski\GATCHINA,
R.~Novotny\GIESSEN,
M.~Ostrick\PI$^,$\footnotemark[1],
L.~Pant\GIESSEN,
M.~Pfeiffer\GIESSEN,
D.~Piontek\HISKP,
\mbox{A.~Roy\GIESSEN,}
A.V.~Sarantsev\HISKP$^,$\GATCHINA,
Ch.~Schmidt\HISKP,
H.~Schmieden\mbox{\PI,}
S.~Shende\KVI,
A.~S\"ule\PI,
V.V.~Sumachev\GATCHINA,
T.~Szczepanek\HISKP,
U.~Thoma\HISKP,
D.~Trnka\GIESSEN,
R.~Varma\GIESSEN,
D.~Walther\HISKP,
Ch.~Wendel\HISKP,
A.~Wilson\HISKP$^,$\FSU\\
 (The CBELSA/TAPS Collaboration)\\\vspace{8mm}
}

\affiliation{\HISKP Helmholtz-Institut f\"ur Strahlen- und Kernphysik, Universit\"at Bonn, Germany}
\affiliation{\GIESSEN II. Physikalisches Institut, Universit\"at Giessen, Germany}
\affiliation{\GATCHINA Petersburg Nuclear Physics Institute, Gatchina, Russia}
\affiliation{\KVI Kernfysisch Versneller Instituut, Groningen, The Netherlands}
\affiliation{\PI Physikalisches Institut, Universit\"at Bonn, Germany}
\affiliation{\FSU Department of Physics, Florida State University, Tallahassee, FL 32306, USA}
\affiliation{\BASEL Physikalisches Institut, Universit\"at Basel, Switzerland}
\date{\today}

\begin{abstract}\noindent
The $N\pi^0\pi^0$ decays of positive-parity $N^*$ and $\Delta^*$ resonances at about 2\,GeV are studied at ELSA by photoproduction of two neutral pions off protons. The data reveal clear evidence for several intermediate resonances: $\Delta(1232)$, $N(1520){3/2^-}$, and $N(1680){5/2^+}$, with spin-parities $J^P=3/2^+$, $3/2^-$, and $5/2^+$. The partial wave analysis (within the Bonn-Gatchina approach) identifies $N(1440)1/2^+$ and the $N(\pi\pi)_{\rm S-wave}$ (abbreviated as $N\sigma$ here) as further isobars, and assigns the final states to the formation of nucleon and $\Delta$ resonances and to non-resonant contributions. We observe the known $\Delta(1232)\pi$ decays of $\Delta(1910)1/2^+$, $\Delta(1920)3/2^+$, $\Delta(1905)5/2^+$, $\Delta(1950)7/2^+$, and of the corresponding spin-parity series in the nucleon sector, $N(1880)1/2^+$, $N(1900)3/2^+$, $N(2000)5/2^+$, and $N(1990)7/2^+$. For the nucleon resonances, these decay modes are reported here for the first time. Further new decay modes proceed via $N(1440)1/2^+\pi$, $N(1520)3/2^-\pi$, $N(1680)5/2^+\pi$, and $N\sigma$. The latter decay modes are observed in the decay of $N^*$ resonances and at most weakly in $\Delta^*$ decays. It is argued that these decay modes provide evidence for a 3-quark nature of $N^*$ resonances rather than a quark-diquark structure. 
\end{abstract}
\maketitle

\noindent

The proposition that mesons and baryons are composed of constituent quarks \cite{GellMann:1964nj,Zweig:1964jf} paved the path to an understanding of the particle zoo. Quark models were developed which reproduced the masses of ground-state baryons and the gross features of their excitation spectrum \cite{Isgur:1978,Glozman:1995fu,Capstick:2000qj,Loring:2001kx}. However, important details remained unexplained like the masses of $N(1440){1/2^+}$ \cite{Krehl:1999km},  $N(1535){1/2^-}$ \cite{Kaiser:1995cy}, and $\Lambda(1405){1/2^-}$ \cite{Jido:2003cb}. These - and other - resonances can be generated dynamically from the interaction between their decay products, and no quark degrees of freedom are required to understand their properties \cite{Kolomeitsev:2003kt}. High-mass baryon excitations fall onto linear Regge trajectories having the same slope as mesonic Regge trajectories. This observation has led to the suggestion that baryon excitations can be interpreted as $q$-$qq$ excitations where the diquark remains in a relative $S$-wave. Here, we use the concept of diquarks in this limited sense even though in the literature, any pair of quarks is also often called a diquark. The quark-diquark picture of baryons entails a much reduced number of expected resonances and may thus explain the small number of actually observed resonances, a fact known as the {\it missing-resonance problem}. Indeed AdS/QCD \cite{Brodsky:2010kn}, based on the string nature of the strong interaction, reproduces the $N^*$ and $\Delta^*$ spectrum very successfully \cite{Forkel:2008un}. Quark models are also incompatible with the existence of {\it spin-parity doublets}, pairs of resonances with similar masses, identical spin and opposite parities. It has been suggested that chiral symmetry might be restored when baryons are excited \cite{Glozman:1999tk,Jaffe:2004ph,Glozman:2007ek,Glozman:2008vg}. One question must find an answer: Is one of these different approaches the correct one? Or, do they just represent different legitimate views?  

The distinctive feature of quark models of baryons as compared to other concepts is the three-body nature of the interaction. In quark models, the internal dynamics is described by two oscillators, usually denoted as $\vec \lambda$ and $\vec\rho$. In the simplest approximation, both oscillators are harmonic; in more realistic cases, mixing of quark model states occurs.  Neglecting mixing, the four positive-parity $\Delta^*$ resonances form a spin-quartet with intrinsic orbital and spin angular momenta $L=2$ and $S=3/2$. Indeed, four isolated positive-parity $\Delta^*$ resonances with similar masses exist: $\Delta(1910)1/2^+$, $\Delta(1920)3/2^+$, $\Delta(1905)5/2^+$, $\Delta(1950)7/2^+$, with $J^P=1/2^+,3/2^+,5/2^+,7/2^+$. Quark models predict an additional $J^P=3/2^+,5/2^+$ spin doublet at about the same mass for which no evidence exists. If the doublet exists, it must have weak coupling to $\pi N$ and  $\gamma N$ and can be neglected in the present discussion.  

Neglecting mixing with the hypothetical $J^P=3/2^+,5/2^+$ spin doublet, the four positive-parity $\Delta^*$ resonances have a spin and a flavor wave function which is symmetric with respect to the exchange of any pair of quarks. The color wave function is completely antisymmetric, hence  the spatial wave function $\phi_{n,l}$ must be symmetric. It can be cast into the form\vspace{-3mm}

     {\footnotesize
\begin{equation}
\label{S}
S= \frac{1}{\sqrt{2}}\bigl\lbrace
    \left[\phi_{0s}(\vec \rho)\times\phi_{0d}(\vec \lambda)\right]
	+
    \left[\phi_{0d}(\vec \rho)\times\phi_{0s}(\vec \lambda)\right]\bigr\rbrace^{(L=2)}.
\end{equation}}
Here, the $\rho$ and the $\lambda$ oscillators are coherently excited from $l_\rho$\,=\,$l_\lambda$\,=\,$0$ ($s$-wave) to $l_\rho$\,=\,$2$ or $l_\lambda$\,=\,$2$ ($d$-wave); the oscillation energy fluctuates from $\rho$ to $\lambda$ and back to the $\rho$ oscillator. There is no radial excitation, $n_\rho$\,=\,$n_\lambda$\,=\,$0$. 

There are as well four isolated positive-parity nucleon resonances: $N(1880)1/2^+$, $N(1900)3/2^+$, $N(2000)5/2^+$, $N(1990)7/2^+$. These resonances have a $2^*$ rating only \cite{Agashe:2014kda}, except for the $3^*$ $N(1900)3/2^+$ \cite{Nikonov:2007br} which is observed in several partial wave analyses \cite{Burkert:2014wea}. Here, we assume that all four resonances exist and that they form a spin-quartet  \cite{Anisovich:2011su}. Nucleon states with $S=3/2$ require spatial wave functions of mixed symmetry.  For $L=2$ the wave functions have equal admixtures of\vspace{-3mm}

  {\footnotesize\begin{eqnarray}
\label{MS}
\mathcal{M_S}
    & = &\frac{1}{\sqrt{2}}\bigl\lbrace
    \left[\phi_{0s}(\vec \rho)\times\phi_{0d}(\vec \lambda)\right]-\left[\phi_{0d}(\vec \rho)\times\phi_{0s}(\vec \lambda)\right]\bigr\rbrace^{(L=2)} \\
\label{MA}\mathcal{M_A} & = &
    \left[\phi_{0p}(\vec \rho)\times\phi_{0p}(\vec \lambda)\right]^{(L=2)}\,,
\end{eqnarray}}
and both parts need to be present to fulfill the Pauli principle. The part $\mathcal{M_A}$ describes a component in which the $\rho$ and the $\lambda$ oscillator are both excited simultaneously. 

The quark model also predicts a spin doublet of positive-parity $N^*$ resonances with $J=1/2$ and $3/2$ and a total intrinsic orbital angular momentum $L=1$. Their spatial wave functions are completely antisymmetric:\vspace{-3mm}

{\footnotesize\begin{eqnarray}
\label{A}
\mathcal{A} & = &
    \left[\phi_{0p}(\vec \rho)\times\phi_{0p}(\vec \lambda)\right]^{(L=1)}\,.
\end{eqnarray}}
In these two $N^*$ resonances, both oscillators are excited. It has been argued \cite{Hey:1982aj}
that these resonances cannot be formed in a $\pi N$ scattering experiment, that it is not possible to excite both oscillators in a single step. If they are produced, they cannot decay directly into $\pi N$. We call this argument the {\it Hey-Kelly conjecture}.

We assume that the type-(\ref{S}) and (\ref{MS}) wave functions can easily disintegrate into $\pi N$ or into $\pi \Delta(1232)$ or other modes with a ground-state baryon and a pseudoscalar (or vector) meson. The {\it Hey-Kelly conjecture} implies that the type-(\ref{MA}) component decays only by consecutive de-excitations of the two oscillators. Hence in a first step, an intermediate resonance with intrinsic orbital angular momentum should be formed, either in the baryon or the meson part. A spin-parity quartet of positive-parity $N^*$ resonances must have a type-(\ref{MA}) component. Hence we expect non-vanishing branching ratios for decay modes like $N(1535)1/2^-\pi$, $N(1520)3/2^-\pi$.    

\begin{figure}[pt]
\includegraphics[width=0.50\textwidth]{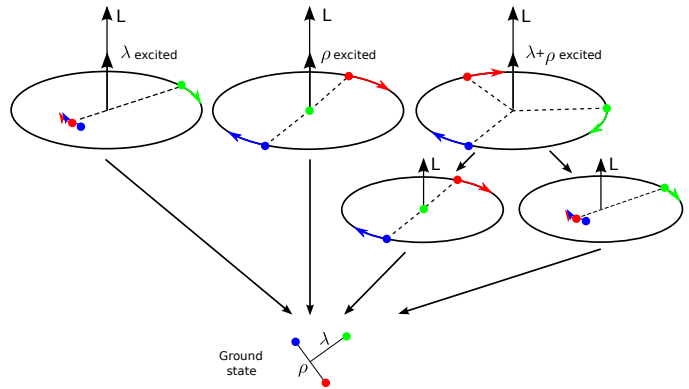}
\caption{\label{fig:cascade}Classical orbits of nucleon excitations with $L=2$ (upper row) and $L=1$ (lower row). The first two pictures in both rows show excitations of the $\rho$ and $\lambda$ oscillators, in the third picture in the first row both, $\rho$ and $\lambda$ are excited. }
\vspace{-3mm}
\end{figure}

The situation is visualized in Fig.~\ref{fig:cascade} using a classical picture. The upper subfigures show the three configurations: (\ref{S}) where the $\rho$ oscillator is in the ground state and the diquark and the third quark rotate around the center of mass; (\ref{MS}) where the two quarks of the diquark carry the angular momentum; and (\ref{MA}) where all three quarks rotate around the center of mass. According to the {\it Hey-Kelly conjecture}, the configuration~(\ref{MA}) does not de-excite directly into a ground-state baryon, $N$ or $\Delta(1232)$, plus a pseudoscalar (or vector) meson: configuration~(\ref{MA}) de-excites into an intermediate state carrying angular momentum in either the $\lambda$ or the $\rho$ oscillator. These decay modes are depicted in the two lower subfigures.

In this paper we report decay modes of $N^*$ and $\Delta^*$ resonances produced in the reaction
${\gamma} p\rightarrow \pi^{0}\pi^{0}{p}$. The results cover $N^*$ and $\Delta^*$ resonances in the mass range from 1500 - 2100\,MeV; analysis and results are documented in detail elsewhere \cite{Sokhoyan:2014tbd}. Here we present only results on positive-parity resonances in the 1900 - 2100\,MeV mass region.

The data were obtained using the energy-tagged photon beam of the ELectron Stretcher Accelerator (ELSA)~\cite{Hillert-EPJA} and the CBELSA/TAPS detector setup \cite{Elsner-EPJA1,Gutz:2014wit}. The electrons hit a bremsstrahlung target and their momenta were analyzed in a dipole magnet in combination with a scintillator hodoscope. The photons then impinged on a 5\,cm long liquid hydrogen target~\cite{Kopf-PhD}, located in the center of the
electromagnetic calorimeter setup consisting of  the Crystal Barrel \cite{Aker-NIM} with 1290 CsI(Tl) crystals and the TAPS detector \cite{Nowotny-IEEE,Gabler-NIM} in a forward wall setup with 528
BaF$_2$ modules. The latter modules are covered with plastic scintillators for charge information. The ca\-lo\-ri\-me\-ters cover the complete azimuthal angle and polar angles from $6^\circ$ to $168^\circ$.
Further information on charged particles is provided by a three-layer scintillating fiber detector~\cite{Suft-NIM} surrounding the target.  After a series of kinematic cuts, the data were subjected to kinematic fits \cite{Pee-EPJA} to the $\gamma p\rightarrow \pi^{0}\pi^{0}p$ hypothesis, and about 1\,600\,000 events in the photon energy range of 900 to 2500\,MeV were retained with a background contamination over the full energy range
of below~1\%. The data were included in the BnGa data base (see \cite{Anisovich:2011fc}
and \cite{Anisovich:2013vpa} for recent additions) covering pion and photon-induced reactions. Data on three-body final states were included in the partial wave analysis in an event-by-event likelihood fit. The fitted mass and angular distributions are compared to data in \cite{Sokhoyan:2014tbd}.

\begin{figure}[pt]
\includegraphics[width=0.48\textwidth,height=0.37\textwidth]{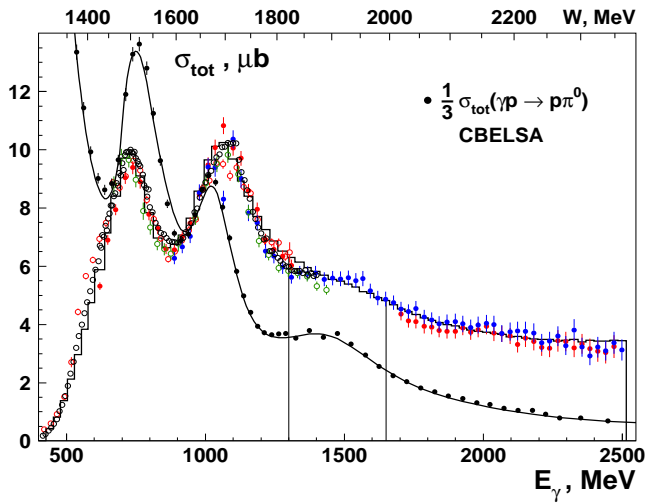}
\caption{\label{FigureTotals}The total cross section for
$\gamma p\to p\pi^0\pi^0$. The red and blue  full dots are from two running periods of this experiment. The red open circles are from CBELSA~\cite{Thoma:2007bm}, green open circles from GRAAL \cite{Assafiri:2003mv}, the black open circles are derived from A2~\cite{Kashevarov:2012wy}, for further details see~\cite{Sokhoyan:2014tbd}. Results on single $\pi^0$ photoproduction \cite{Bartholomy:2004uz} are shown for comparison. The results of the partial wave analysis are shown as solid lines, for the $p\pi^0$ channel scaled down by a factor 3. \vspace{-2mm} }
\end{figure}

\begin{figure}[pt]\vspace{-5mm}
\includegraphics[width=0.46\textwidth]{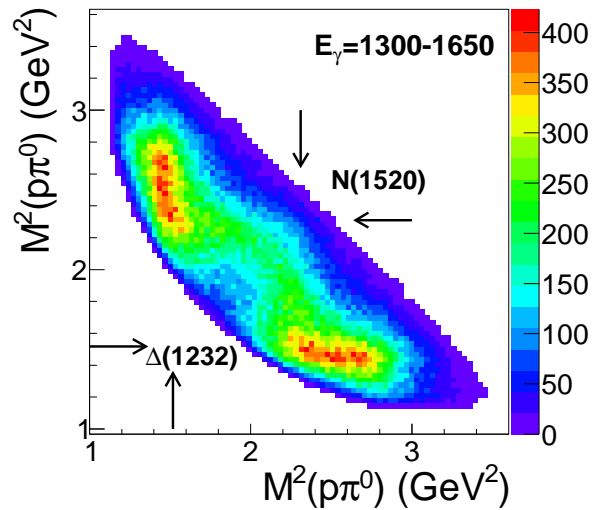}\vspace{-2mm}
\caption{\label{fig:dalitz}The $p\pi^0\pi^0$ Dalitz plot for $1300<E_\gamma<1650$\,MeV covering the fourth resonance region. There are two entries per event. The largest contributions are due to cascade processes with $\Delta(1232)$ as intermediate state. Two further bands are seen due to cascades via $N(1520)$ formation. In regions I and II, the fractional contributions of $\Delta(1232)$ and $N(1520)$ are enhanced. The depletion of the Dalitz plot at its borders is due to the wide window in photon energy.\vspace{-2mm} }
\end{figure}

The total cross section for two-neutral-pion photoproduction (Fig.~\ref{FigureTotals}) shows very significant peaks in the 2$^{\rm nd}$  and 3$^{\rm rd}$ resonances regions and a small shoulder at an invariant mass of about $W=1.9$\,GeV  due to the 4$^{\rm th}$   resonance region. For comparison we also show the total cross section for $\gamma p\to \pi^0 p$ from \cite{Bartholomy:2004uz} with the tail of $\Delta(1232)$ and a more visible fourth resonance region.

Figure~\ref{fig:dalitz} shows the $p\pi^0\pi^0$ Dalitz plot
\cite{Sokhoyan:2014tbd} for the range of the proton-$\gamma$ invariant
mass from $W=1820$ to $2000$\,MeV in the . Since the two neutral pions are identical, there are two
entries per event, and the Dalitz plot is symmetric with respect to
the diagonal. The largest contributions are seen at squared $p\pi^0$
invariant masses around 1.5\,GeV$^2$ stemming from the
$\Delta(1232)$ resonance as intermediate state. A smaller band can
be seen at $M^2\approx 2.25\, \rm GeV^2$ or $M\approx 1500$\,MeV. A
fit returns mass and width compatible with the $N(1520)3/2^-$ resonance.

\begin{table*}
\caption{\label{tab:DeltaN}Branching ratios (in \%) for the decays of nucleon and $\Delta$ resonances. The errors are derived from a large number of fits. $\times$ stands for forbidden, $-$ for allowed decay modes which in all fits converged to zero. Further properties of these (and other) resonances are reported in \cite{Sokhoyan:2014tbd}. } 
\begin{center}
\renewcommand{\arraystretch}{1.4}    
{\scriptsize
\begin{tabular}{||l||cc|cc|cc||c|c|cc||c|l||}
\hline\hline
                   &$N\pi$&$L$  &$\Delta\pi$ &\hspace{-3mm}$L$$<$$J$\hspace{-3mm}&$\Delta\pi$ &\hspace{-3mm}$L$$>$$J$\hspace{-3mm}&$N(1520)\pi$ $L$&$N(1535)\pi$ $L$&$N\sigma$&$L$&$N(1440)\pi$ $L$ &$N(1680)\pi$ $L$\\ \hline
$\Delta(1910)1/2^+$&12\er 3   &1&$\times$&&50\er 16&1&\quad-\hfill 0&5\er 3\hfill 2           &$\times$&&6\er 3\hfill 1&\quad -\hfill 3\\
$\Delta(1920)3/2^+$& 8\er 4   &1&18\er 10&1&58\er14&3& $<5$\hfill 0&$<2$\hfill 2         &$\times$&&$<4$\hfill 1&\quad -\hfill 1\\
$\Delta(1905)5/2^+$&13\er 2   &3&33\er10&1&-&3&\quad -\hfill 2 &$<1$\hfill 2         &$\times$&&\quad -\hfill 3 &10\er5\hfill 1\\
$\Delta(1950)7/2^+$&46\er 2   &3&5\er 4&3&-&5&\quad-\hfill 2&\quad-\hfill 4       &$\times$&&\quad-\hfill 3& \ 6\er3\hfill 1\\[0.2ex]
\hline\hline\\[-3.5ex]
$N(1880)1/2^+$     & 6\er 3   &1&$\times$&&30\er 12&1&\quad-\hfill 2&8\er 4 \hfill 0          &25\er 15&0&\quad-\hfill 1&\quad -\hfill 3\\
$N(1900)3/2^+$     & 3\er 2   &1&17\er 8&1&33\er 12&3&15\er 8\hfill 0&7\er 3\hfill 2    &4\er 3&2&$<$2 \hfill 1&\quad -\hfill 1\\
$N(2000)5/2^+$     & 8\er 4   &3&22\er 10&1&34\er15&3&21\er 10\hfill 2&\quad-\hfill 2  & 10\er 5&2&\quad-\hfill 1&16\er9 \hfill 1\\
$N(1990)7/2^+$     & 1.5\er0.5&3&48\er 10&3&-&5&$<$2\hfill 2&$<$2\hfill 4                &-&4&$<$2\hfill 1&\quad -\hfill 1\\
\hline
\hline
\end{tabular}
\renewcommand{\arraystretch}{1.0}}
\end{center}
\vspace{-3mm}
 \end{table*}
The results on observed decay modes of positive-parity resonances in the 1850 to 2000\,MeV region are collected in Table~\ref{tab:DeltaN}. The table includes results on $N(1535)1/2^-\pi$ decay modes from \cite{Gutz:2014wit}. All resonances have sizable $N\pi$ and/or $\Delta\pi$ branching ratios: for $\Delta^*$ resonances, the sum of the two decay branching ratios is 61\% on average, for $N^*$'s it is 51\%. There are several allowed branching ratios into intermediate resonances carrying one unit of intrinsic orbital angular-momentum excitation ($N(1520)3/2^-\pi$, $N(1535)1/2^-\pi$, $N\sigma$). Their mean contribution to the four $N^*$'s is 23\%, and to the four $\Delta^*$'s a factor 10 smaller. In the $N\sigma$ decay mode, we assume that first, a $q\bar q$ pair is created, the $q$ forms - together with the de-excited $qq$ pair, the final state baryon, and the $\bar q$ picks up the third quark of the primary baryon resonance (which still carries angular momentum) and produces a $q \bar q$ pair with $J^P=0^+$ which dresses to become a $\sigma$.

The $N(1440)1/2^+\pi$ and $N(1680)5/2^+\pi$ branching ratios are also given even though the intermediate state does not carry one unit of orbital angular momentum as Fig.~\ref{fig:cascade} seems to imply. We assume that in the de-excitation process, part of the excitation energy can be transferred to an oscillator in a single step transition. In atomic physics this is known as Auger effect. 

The individual branching ratios have large errors bars. Hence we fitted the data with two assumptions: 
i) we forbade decays of the four $\Delta^*$'s into  $N(1440)\pi$, $N(1520)\pi$, $N(1535)\pi$, and 
$N(1680)\pi$. This has little effect on the fit and the $\chi^2$ deteriorated by 692 units. We 
consider this to be at the border of becoming statistically significant. Of course, these decays 
are not forbidden but obviously, the branching ratios for these decay modes from the four 
$\Delta^*$'s are small. 
 ii) if decay modes into orbitally excited states were forbidden for the four $N^*$ resonances, the 
$\chi^2$ change became 3880 units and the fit quality deteriorated visibly. The four $N^*$  resonances 
decay via orbitally excited intermediate states with a significant decay fraction.

Why are the decays of the four $N^*$ resonances into orbitally excited intermediate resonances so frequent ($\approx 23$\%), and why are these decay modes suppressed for the four $\Delta^*$ resonances? We assign the difference to the component type~(\ref{MA}) in the wave function of the four $N^*$'s which is absent in the  wave function of the four $\Delta^*$'s. The type~(\ref{S}) and type~(\ref{MS}) components disintegrate easily into $\pi N$ or $\pi\Delta(1232)$, a decay mode which is -- according to the {\it Hey-Kelly conjecture} -- suppressed from the type~(\ref{MA}) component.  The decays of the four $N^*$ resonances into orbitally excited intermediate resonances thus provide evidence for a three-body component in their wave functions and for a significant coupling of this component to orbitally excited intermediate resonances.

Summarizing, we have reported a study of photoproduction of two neutral pions off protons in a range of photon energies covering the fourth resonance region and have identified several new decay modes of known $N^*$ and $\Delta^*$ resonances. The $N^*$ resonances have non-vanishing branching ratios into the excited $N^*$ resonances $N(1520)3/2^-$ and $N(1535)1/2^-$ while for the $\Delta^*$ these decay modes are suppressed. The pattern suggests that $N^*$ resonances in the fourth resonance region contain a sizable component in their wave function in which two oscillators are excited. This observation supports interpretations of baryon resonances exploiting the full three-body dynamics and challenges models assuming a quark-diquark structure. 

We thank the technical staff of ELSA and the par\-ti\-ci\-pating institutions for their invaluable contributions. We acknowledge support from the \textit{Deutsche Forschungsgemeinschaft} (SFB/TR16), \textit{Schweizerischer Nationalfonds}, and \textit{U.S. National Science Foundation}.


\begin{thebibliography}{99}
\bibitem{GellMann:1964nj}
  M.~Gell-Mann,
  Phys.\ Lett.\  {\bf 8}, 214 (1964).
\bibitem{Zweig:1964jf}
  G.~Zweig,
``An SU(3) model for strong interaction symmetry and its
breaking.,'' in: 'Developments in the Quark Theory of
Hadrons'. D. Lichtenberg and S. Rosen (eds.). Nonantum,
Mass., Hadronic Press (1980) 22-101.
\bibitem{Isgur:1978}
  N.~Isgur, G.~Karl,
  Phys.\ Rev.\ D {\bf 18}, 4187 (1978).
\bibitem{Glozman:1995fu} 
  L.Y.\,Glozman,\,D.O.\,Riska,
\,Phys.\,Rept.\,{\bf 268},\,263\,(1996).
\bibitem{Capstick:2000qj} 
  S.~Capstick, W.~Roberts, Prog. Part. Nucl. Phys. {\bf 45}, S241 (2000).	
\bibitem{Loring:2001kx}
  U.~L\"oring {\it et al.},
  Eur.\ Phys.\ J.\ A {\bf 10}, 395 (2001).
\bibitem{Krehl:1999km}
O. Krehl {\it et al.}, 
Phys. Rev. C {\bf 62}, 025207 (2000).
\bibitem{Kaiser:1995cy}
  N.~Kaiser {\it et al.}, 
  Phys.\ Lett.\  B {\bf 362}, 23 (1995).
\bibitem{Jido:2003cb}
  D.~Jido {\it et al.}, 
  Nucl.\ Phys.\  A {\bf 725}, 181 (2003).
\bibitem{Kolomeitsev:2003kt} 
  E.~Kolomeitsev, M.~Lutz,
  Phys.\ Lett.\ B {\bf 585}, 243 (2004).
	\bibitem{Brodsky:2010kn} See, e.g,  S.J.~Brodsky, F.~Guy de
Teramond,
  Chin.\ Phys.\  C {\bf 34}, 1 (2010).
\bibitem{Forkel:2008un}
  H.~Forkel, E.~Klempt,
  Phys.\ Lett.\  B {\bf 679}, 77 (2009).
\bibitem{Glozman:1999tk}
  L.Y.~Glozman,
  Phys.\ Lett.\  B {\bf 475}, 329 (2000).
\bibitem{Jaffe:2004ph}
  R.L.~Jaffe,
  Phys.\ Rept.\  {\bf 409}, 1 (2005).
\bibitem{Glozman:2007ek}
  L.Y.~Glozman,
  Phys.\ Rept.\  {\bf 444}, 1 (2007).
\bibitem{Glozman:2008vg}
  L.Y.\,Glozman,\,A.V.\,Nefediev,\,Nucl.\,Phys.\,A\,{\bf 807},\,38\,(2008).
 \bibitem{Anisovich:2011su} 
  A.V.~Anisovich {\it et al.}, 
  Phys.\ Lett.\ B {\bf 711}, 167 (2012).
	\bibitem{Agashe:2014kda} 
  K.A.~Olive {\it et al.},
  Chin.\ Phys.\ C {\bf 38}, 090001 (2014).
	\bibitem{Nikonov:2007br} 
  V.A.~Nikonov {\it et al.}, 
  Phys.\ Lett.\ B {\bf 662}, 245 (2008).
	\bibitem{Burkert:2014wea} 
  V.D.~Burkert 
	{\it et al.},
  arXiv:1412.0241 [nucl-ex].
\bibitem{Hey:1982aj} 
  A.J.G.~Hey and R.L.~Kelly,
  Phys.\ Rept.\  {\bf 96}, 71 (1983). 
	\bibitem{Sokhoyan:2014tbd}
	V.~Sokhoyan {\it et al.}, ``High statistics study of the reaction $\gamma p\to p\;2\pi^0$, in preparation. 
	\bibitem{Hillert-EPJA}
W.~Hillert, Eur. Phys. J. A \textbf{28}, 139 (2006).
\bibitem{Elsner-EPJA1}
D.~Elsner {\it et al.}, Eur. Phys. J. A \textbf{33} (2), 147 (2007).
\bibitem{Gutz:2014wit} 
  E.~Gutz {\it et al.},
 Eur.\ Phys.\ J.\ A {\bf 50}, 74 (2014).
\bibitem{Kopf-PhD}
B.~Kopf, Ph.D. thesis, Dresden (2002).
\bibitem{Aker-NIM}
E.~Aker {\it et al.}, Nucl. Inst. and Meth. A  \textbf{321}, 69
(1992).
\bibitem{Nowotny-IEEE}
R.~Novotny, IEEE Trans. Nucl. Sci. \textbf{NS-38}, 379 (1991).
\bibitem{Gabler-NIM}
A.R.\,Gabler\,{\it et\,al.},\,Nucl.\,Inst.\,\&\,Meth.\,A\,\textbf{346},\,168\,(1994).
\bibitem{Suft-NIM}
G.\,Suft {\it et al.}, Nucl. Inst. \& Meth. A \textbf{538}, 416 (2005).
\bibitem{Pee-EPJA}
H.~van Pee {\it et al.} Eur. Phys. J. A \textbf{31}, 61 (2007).
\bibitem{Anisovich:2011fc} 
  A.V.~Anisovich {\it et al.},
  Eur.\ Phys.\ J.\ A {\bf 48}, 15 (2012).
	\bibitem{Anisovich:2013vpa} 
  A.V.~Anisovich {\it et al.},
  Eur.\ Phys.\ J.\ A {\bf 49}, 158 (2013).
\bibitem{Thoma:2007bm} 
  U.~Thoma {\it et al.},
  Phys.\ Lett.\ B {\bf 659}, 87 (2008).
\bibitem{Assafiri:2003mv} 
  Y.~Assafiri {\it et al.},
  Phys.\ Rev.\ Lett.\  {\bf 90}, 222001 (2003).
%
\bibitem{Kashevarov:2012wy}
  V.L.~Kashevarov {\it et al.},
  Phys.\ Rev.\ C {\bf 85}, 064610 (2012).
\bibitem{Bartholomy:2004uz} 
  O.~Bartholomy {\it et al.},
  Phys.\ Rev.\ Lett.\  {\bf 94}, 012003 (2005).
\end{thebibliography}
\end{document}